\documentclass[12pt]{iopart}
\usepackage[latin1]{inputenc}
\usepackage{graphicx}
\usepackage{iopams} 
\usepackage{color}
\usepackage{url}
\usepackage{verbatim}

\begin{document}

\title[Investigation of slow collisions for (quasi) symmetric heavy systems]{Investigation of slow collisions for (quasi) symmetric heavy systems: what can be extracted from high resolution X-ray spectra}

\author{M. Trassinelli, C. Prigent, E. Lamour, F. Mezdari\footnote{Present addres: Departement of Physics, Faculty of Sciences of Gabes Cité Erriadh 6072, Zrig, Gabes, Tunisia.}, J. Mérot, R. Reuschl\footnote{Present address: ExtreMe Matter Institute EMMI, GSI Helmholtzzentrum, 64291 Darmstadt, Germany.}, J.-P. Rozet, S. Steydli and D. Vernhet}

\address{CNRS, INSP, UMR7588, 4 Place Jussieu, 75005 Paris, France\\
Université Pierre et Marie Curie, INSP, UMR7588, 4 Place Jussieu, 75005 Paris, France}
\ead{martino.trassinelli@insp.jussieu.fr}
\begin{abstract}
We present a new experiment on (quasi) symmetric collision systems at low-velocity, namely Ar$^{17+}$ ions ($v=0.53$~a.u.) on gaseous Ar and N$_2$ targets, using low- and high-resolution X-ray spectroscopy. Thanks to an accurate efficiency calibration of the spectrometers, we extract absolute X-ray emission cross sections combining low-resolution X-ray spectroscopy and a complete determination of the ion beam - gas jet target overlap. Values with improved uncertainty are found in agreement with previous results \cite{Tawara2001}. Resolving the whole He-like Ar$^{16+}$ Lyman series from $n=2$ to 10 with our crystal spectrometer enables to determine precisely the distribution $\{\mathcal{P}_n\}$ of the electron capture probability and the preferential $n_{pref}$ level of the selective single-electron capture. 
Evaluation of cross sections for this process as well as for the contribution of multiple-capture is carried out. Their sensitivity to the $\ell$-distribution of $n$ levels populated by single-electron capture is clearly demonstrated, 
providing a stringent benchmark for theories. 
In addition, the hardness ratio is extracted and the influence of the decay of the metastable $1s2s\ ^3\!S_1$ state on this ratio is discussed.
\end{abstract}
\date{today}

\submitto{\JPB}
\maketitle

\section{Introduction}
The interaction between low velocity ions and atoms or molecules has been extensively investigated in the past decades.
At the low velocity regime (0.4--1 atomic units (a.u.)), the dominant process is electron capture that occurs in selective projectile excited states, the collisional system behaving like a quasi-molecule. It leads to an energy gain of the projectile ion and the populated excited states decay by emission of photons and/or electrons, both carrying information on the collision dynamics. 
During the 1980s, many experiments have been performed with light ions, like C, N, Ne,\ldots interacting with light targets as H$_2$ and He leading to a quite complete understanding of the mechanisms involved in the collision. 
Besides total cross sections determination, the $n$- and $\ell$ population distributions ($\{\mathcal{P}_n\}$ and $\{\mathcal{P}_\ell\}$, respectively) of the single capture process have been characterized as well as the role of rotational versus radial couplings while for double-capture, the $n \ell n' \ell'$ channels have been identified.
For such studies, different detection techniques have been applied, as the kinetic energy gain of the projectile (\cite{Barat1992} and reference therein) or the electron and the radiative emission detection \cite{Mack1989,Vernhet1988,Chetioui1990,Stolterfoht1997,Currel2003}.
At the same time, extended theoretical investigations have been developed.
They include simple approaches such as the well-known Classical Over-Barrier (COB) model or the reaction window within the Landau-Zener model \cite{Ryufuku1980,Taulbjerg1986}, and more sophisticate couple state calculations using basis of either atomic or molecular orbitals \cite{Fritsch1984,Winter1984, Kimura1987,Errea1987}.
With the advent of new ion sources, heavier projectile ions of higher charge states have become available. 
New techniques like the powerful cold-target recoil-ion momentum spectroscopy (COLTRIMS) were applied allowing to get unambiguous determination of partial $n$ single-electron capture including differential cross sections for the main capture channel like in the case of Ar$^{(15-18)+}$ on He at 14~keV q$^{-1}$ for example \cite{Knoop2008}.

However, today, there is still a lack of data for (quasi) symmetric collisions, i.e., highly charged ions colliding on heavy targets, for which contribution from multi-electron exchange processes is important.
The only systematic studies available are restricted to coincidence measurements between charge selected projectile and target ions \cite{Barany1985,Ali1994} where stabilized charge exchange mechanisms are investigated.
More recently, low-resolution X-ray detection experiments have been performed \cite{Beiersdorfer2000,Tawara2001,Tawara2006,Allen2008}. 
X-ray measurements provide data in well defined conditions (in the laboratory frame) that are of interest to interpret X-ray spectra observed from different environments such as those from solar wind colliding with comets \cite{Cravens2002,Krasnopolsky2004}.
However, in both cases, no information on the partial $n \ell$ electron capture states is accessible. Furthermore, in the case of low-resolution X-ray measurements, as discussed recently in \cite{Otranto2011}, the effect of single and multiple processes cannot be distinguished.
Finally, all the laboratory data are used to extract a parameter called the ``hardness ratio'' (i.e., the ratio between the X-ray intensity of $n>2 \to 1$ and $2 \to 1$ transitions) that serves as a reference to determine abundance of elements when interpreting astrophysical X-ray spectra.

With the experiment described in this paper we are bringing new information on the collision dynamics for (quasi) symmetric systems, studying the collisions of low-velocity (0.53~a.u.) Ar$^{17+}$ ions with a gaseous Ar and N$_2$ targets by coupling a complete characterization of the beam-target overlap with low- but also high-resolution X-ray spectroscopy measurements.
On the one hand, low-resolution X-ray spectroscopy allows us for a measurement of the absolute X-ray emission cross-section. 
In addition, by resolving the whole Lyman series $1snp \to 1s^2$ up to $n=10$, we are able to determine the preferential level $n_{pref}$ and $n$-distribution $\{\mathcal{P}_n\}$ of the selective-state single-electron capture. Thanks to the accurate knowledge of our detection spectrometers, the single-electron capture cross section   in highly excited levels has also been extracted.
Moreover, contribution of multiple processes giving rise to autoionizing multiple excited states (i.e., participating to the X-ray lines for states $< n_{pref}$) has also been evaluated. The limitation in determining such cross sections is discussed as well as the relevance of measured hardness ratio with low-energy resolution X-ray spectra in ion-atom collision at a few keV/u.

The article is organized as follows. 
Section~\ref{sec:setup} is dedicated to the description of the general set-up and the X-ray detection system.
In section~\ref{sec:low-resolution}, the method used to extract the absolute value of the X-ray emission cross section is detailed, including the calculation of the beam-target overlap.
In section~\ref{sec:high-resolution}, with the help of the high-resolution spectroscopy, results on the relative population of He-like $1snp$ states for $2\leq n \leq 10$ are presented as well as the determination of $\{\mathcal{P}_n\}$ of single-electron capture. In particular the influence of the $\ell$-sublevels population on the recorded spectra is considered and discussed. Evaluation of single-electron capture cross section and contribution of multiple capture to the X-ray spectra are given. Finally a detailed analysis of the $n = 2 \to 1$ relative intensities observed with the high resolution X-ray spectrometer is described and the significance of the measured hardness ratio is studied.
General remarks and comments are given as conclusions in the last section.

\section{Experimental set-up and detection system} \label{sec:setup}
\subsection{Description of the set-up}\label{sec:setup-det}
\begin{figure}
\begin{center}
\includegraphics[width=0.7\textwidth]{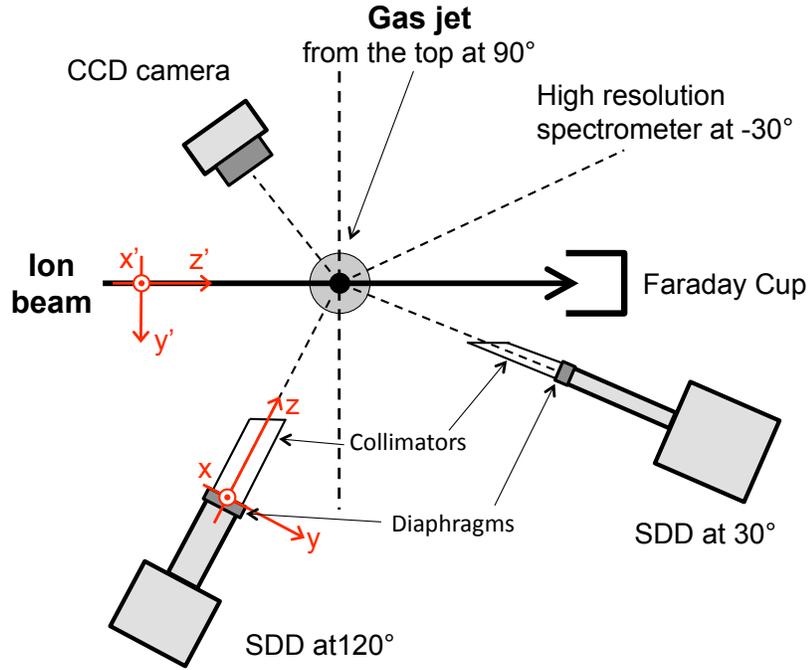}
\caption{(color online) A schematic view of the experimental set-up showing the localization of the two solid-state detectors, the high resolution X-ray spectrometer, the CCD camera and the gaseous jet target.} 
\label{fig:setup}
\end{center}
\end{figure}

\begin{figure}
\begin{center}
\includegraphics[width=0.55\textwidth]{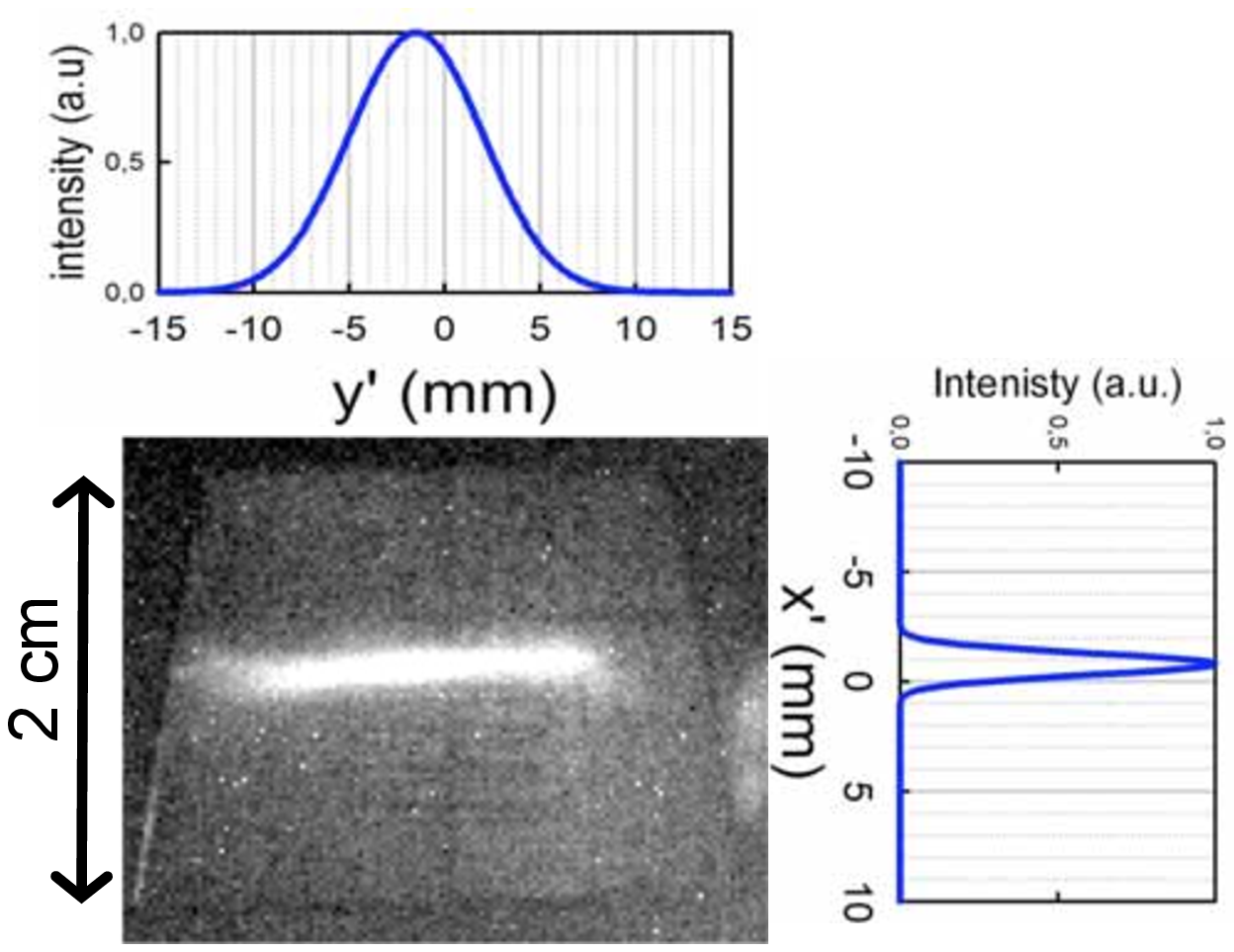}
\includegraphics[width=0.4\textwidth]{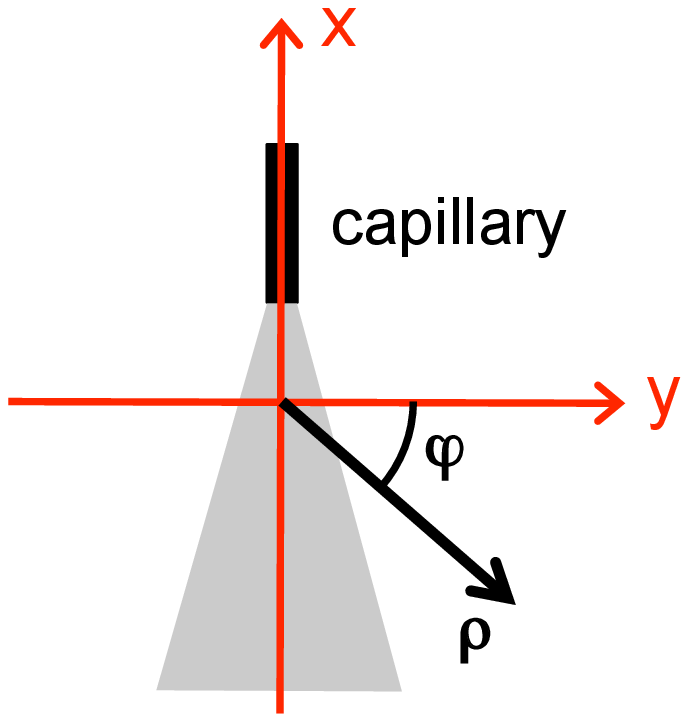}
\caption{(color online) Left: image of the transversal ion beam profile from a CCD camera obtained by the impact of the incoming argon ions on an stainless steel target. Right: target density profile with the coordinate system used for the calculation of the atomic density of the effusive gas jet. } 
\label{fig:profiles}
\end{center}
\end{figure}

The experiment has been performed at the low energy ion installation ARIBE\footnote{French acronym for ``Accelerator dedicated to research with low energy ions''.}\cite{Maunoury2002} at GANIL\footnote{Grand Acc\'el\'erateur National d'Ions Lourds, Caen, France.} where 255~keV $^{36}$Ar$^{17+}$ ions are directed onto a gaseous jet of neutral Ar and N$_2$.
The optic equipment of the ion beam is described in \cite{Prigent2009}.
A schematic view of the experimental set-up is presented in figure~\ref{fig:setup}, showing the arrangement of the X-ray spectrometers, the Charge-Coupled Device (CCD) camera and the gaseous jet target around the interaction zone.  
Lyman lines (in the 3.1--4.1~keV energy range) from the ion projectile are recorded by two Silicon Drift Detectors (SDD) placed respectively at $+30^\circ$, $+120^\circ$ from the ion beam axis, and by a high-resolution high-transmission Bragg crystal spectrometer located at $-30^\circ$.
To maximize the resolving power of our crystal spectrometer, the ion beam is vertically focused to 1--3~mm and limited to a beam width of 10--15~mm by reducing the beam emittance.
The beam profile is regularly monitored detecting, with a high-sensitivity CCD camera, the fluorescence light from the ion impact on a stainless steel or alumina target (see figure~\ref{fig:profiles} (left) as an example). 
A Faraday cup installed after the crossing zone and coupled to a current integrator, measures continuously the beam current. 
A maximum intensity of 2~enA was obtained.

The gaseous jet target crosses the ion beam at $90^\circ$ and is obtained by gas diffusion through a capillary into the interaction chamber. Of effusive type, this gas source, widely used in several previous experiments, has been well characterized 
\cite{Giordmaine1960,Gray1992,Rugamas2000}. 
Atomic density and flow are determined by the geometry of the jet, i.e., the length $L$ and the diameter $d$ of the capillary, and by the thermodynamics conditions, i.e., the temperature and the backing pressure $p$ of the gas. 
In our case, the characteristic dimensions are $L=35$~mm, $d=0.1$~mm, and working at room temperature, $p=0.5-5$~mbar.
These conditions correspond to the so-called ``opaque intermediate'' regime, for which the typical dimensions of the capillary are comparable to the gas mean free path and, more specifically, the Knudsen numbers are $K_L\lesssim 1$ and $K_d<1$ \cite{Giordmaine1960,Gray1992}. 
In this case inter-atomic/molecular and wall-atom/molecule collisions inside the capillary play a role on the density profile at the exit of the capillary resulting in the so-called intermediate regime between the viscous and molecular flow.
For a backing pressure $p=1.5$~mbar, it leads, at the crossing point with the ion beam (i.e., 6 mm from the capillary exit), to an atomic density of about $5 \cdot 10^{12} $ at./cm$^3$ and a spatial extension of 8~mm (corresponding to an angular opening of 35$^\circ$).

Finally it is worth mentioning that the residual gas pressure has an influence on the detected X-ray signal through the production of metastable states in the ion beam interacting with the gaseous jet. Two contributions have to be taken into account, namely the residual vacuum in the beam line, $10^{-7}$~mbar over a distance of 350~cm (the distance between the last dipole and the entrance of the interaction chamber) and within the interaction chamber itself, $10^{-6}$~mbar over 20~cm (the radius of the interaction chamber) when the gaseous target is ``on''. 

\subsection{X-ray spectrometers and experimental conditions}\label{sec:detectors}
The two SDDs (XFlash model detector from ROENTEC and AXAS model detector from KETEK), having been extensively studied in terms of efficiency over their whole range of energy detection \cite{Lamour2009}, are used to determine the absolute X-ray emission cross sections.  
Their efficiency (including the quantum efficiency and the window transmission) in the energy range of interest is close to one. They are respectively $(91.70 \pm 0.35)\%$ and $(86.6  \pm 1.0) \%$ at 3~keV.
They are placed at an observation angle of 30$^\circ$ and 120$^\circ$ with respect to the ion beam, and are equipped with long specifically designed collimators in which a diaphragm is inserted for background reduction purpose (see figure~\ref{fig:setup}). 
This collimation system gives rise to a solid angle of  $8 \cdot 10^{-3}$~sr and  $1.6 \cdot 10^{-3}$~sr for the detector placed at  30$^\circ$ and 120$^\circ$, respectively.
These observation angles have been chosen, on the one hand, to monitor the crystal spectrometer (the SDD at 30$^\circ$, see next \textsection) and, on the other hand, at an angle to be nearly insensitive to the possible polarization of the emitted X-rays (120$^\circ$ close to the magic angle).
The SDDs record the complete series of Lyman transitions with an energy resolution of typically 190~eV and 210~eV at 3~keV for the SDDs placed at 30$^\circ$ and 120$^\circ$, respectively.

A Bragg crystal spectrometer is used to reach a much higher resolution. 
Details about its principle of operation are given in \cite{Rozet1985}.
Briefly summarized here, high transmission is achieved by using a Highly Oriented Pyrolitic Graphite (HOPG) crystal with a mosaic spread of 0.4$^\circ$, and a large surface ($60 \times 60$ mm$^2$) position-sensitive multi-wire gas-detector. The spectrometer is used in a vertical geometry and placed at an angle of 30$^\circ$ in order to be insensitive to the possible polarization of the recorded 3.1 to 4.1~keV line intensities (the crystal itself acting as a polarimeter \cite{Vernhet1998}).  
In addition, setting the target-to-crystal and crystal-to-detector distances at an equal value $D$ allows removing, at first order, the broadening associated to the mosaic spread of the crystal. A choice of $D= 800$~mm corresponds to an optimum in both energy resolution and transmission efficiency \cite{Gumberidze2010}. Under these conditions, the largest remaining contribution to the resolution comes from the beam height, which has been limited to $1-3$~mm (as mentioned above), while the spatial resolution of the position sensitive detector, is better than 500 $\mu$m. Finally, depending on the Bragg angle value chosen for the crystal, a resolution power of about 1000 has been achieved at 3.1~keV (i.e., a resolution of 3.2~eV), and better than 500 on an energy range around 4~keV (i.e., a resolution of around 7.2~eV).
Typically, the accessible range at a given setting is about 160~eV.
Consequently all the He-like $1snp \to 1s^2$ transitions generated during the collision are resolved as well as the fine structure of $n=2\to1$ transitions from He- and Li-like ions. 
The detection efficiency ratio between the crystal spectrometer and the SDD, symmetrically positioned with regard to the beam axis, has been precisely determined and it is $0.104 \pm 0.002$ for 3.1~keV X-rays. 

During the data acquisition background measurements are performed with the gas jet ``off'', and with or without the ion beam in the chamber. In the case of the SDD the major source of background is due to the interaction between the Ar$^{17+}$ ion beam and the residual vacuum in the chamber. A typical background spectrum gives rise in fact to a physical signal very similar to the X-ray Lyman lines recorded with the gas jet ``on'' but its contribution is reduced to around 2\%. In the case of the crystal spectrometer, on the contrary, the background is mainly due to the hard X-ray residual radiation produced by the ion source. It leads to a smoothed shape spectrum whatever the Bragg angle chosen and is less than 2 counts/s integrated over the corresponding energy range; its contribution to the different lines observed will be discussed in section 4.  
The data were pre-analyzed and recorded with a GANIL-type acquisition\footnote{\url{http://wiki.ganil.fr/gap}.}   consisting in a series of NIM and VME modules coupled to a computer. The dead-time of the acquisition was constantly monitored by a pulse signal of well defined rate connected to the pre-amplifier of the detectors.
Additional details of the experimental set-up and the acquisition system can be found in \cite{Prigent2009,Prigent2005}.

\section{Absolute X-ray emission cross sections from low-resolution spectra} \label{sec:low-resolution}
\begin{figure}
\begin{center}
\includegraphics[width=0.48\textwidth]{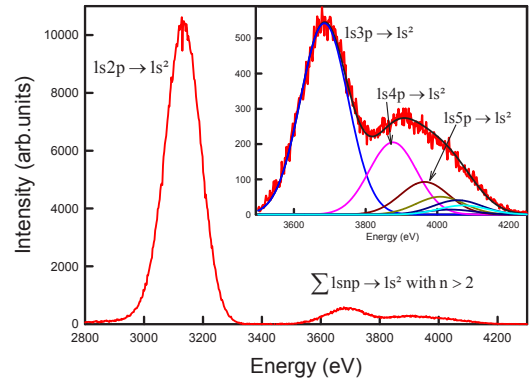}
\caption{(color online) Typical spectrum (with background subtracted) recorded by one of the solid-state detectors exhibiting transitions from Ar$^{16+}$ ions produced after collision with the argon gaseous target ($p=1.5$~mbar).
Experimental data are in red and the fit result is represented by the solid black line. 
For sake of clarity, the different $n>3  \to 1$ transition contributions are shown (solid colored lines) in the zoom inset; their relative intensities are deduced from the analysis of the high-resolution spectra (section \ref{sec:high-resolution}).}
\label{fig:spectra_ion-gas-lr}
\end{center}
\end{figure}

A typical X-ray spectrum from the SDD is presented in figure \ref{fig:spectra_ion-gas-lr}.
Two major peaks are clearly visible in the energy region from 3 to 3.8~keV. As in previous work \cite{Tawara2001,Allen2008} they are assigned to the $1s2p \to 1s^2$ (high intensity line) and $1s3p \to 1s^2$ transitions. The broad peak at higher energy is due to $1snp \to 1s^2$ transitions with $n >3$. 

Quantitative peak intensities have been extracted by fitting the observed lines with a series of Gaussian profiles convoluted to an error function corresponding to the charge collection loss in the SDD detectors although its contribution is relatively small in the $3-4$~keV energy region. For the extraction of the absolute value of the X-ray emission cross section, the ion beam and the gas-jet profiles have to be considered to evaluate precisely the target-projectile overlap.

\subsection{Calculation of the overlap between ion beam and gas-jet target}

We based our calculation on the method developed in ref.~\cite{Wohrer2000}.
In our case we consider the coordinate systems presented in figs.~\ref{fig:setup} and \ref{fig:profiles} where $z$ is the distance between the considered collision point and the SDD detector, and where $\rho$ and $\varphi$ are the radial and angular coordinates with respect to the crossing point.
The overlap integral convoluted with the geometrical efficiency of the detector can be written as follows:
\begin{equation}
B(P)=\int\int\int n_{at}(p,\rho,\varphi)\ \Phi(\rho,\varphi)\ g(\rho,\varphi)\ \frac {\omega(\rho,\varphi,z)} {4 \pi } \rho\ d\rho\ d\varphi\ dz.
\end{equation}
The different terms are described below.

$n_{at}$, the atomic density of the jet, can be written 
\begin{equation}
n_{at}(p,\rho,\varphi)= \frac 1 {\bar v\ \rho^2} \frac {dN} {d\Omega\ dt}(p,\rho,\varphi),
\end{equation}
where $\bar v$ is the arithmetical average of the gas velocity and $dN(\rho,\varphi)/(d\Omega\ dt)$ is the flow for the opaque intermediate regime \cite{Gray1992}. 
$\Phi(\rho,\varphi)$ corresponds to the ion beam profile in the chosen coordinate, assumed to be a Gaussian distribution along the two dimensions (x'-y' in figs.~\ref{fig:setup} and \ref{fig:profiles}) perpendicular to the ion beam propagation.
The third term $g$ corresponds to the geometrical detection efficiency. To properly take into account the shadow-effect induced by the collimator aperture coupled to the diaphragm placed in front of each detector, determination of the geometrical efficiency requires a complete calculation that has been based on a procedure described in refs.~\cite{Gigante1989,Conway2010}.
Finally, $\omega$ is the solid angle subtended by the detector at given coordinates $(\rho,\varphi,z)$.

The number of X-rays, $N_X$, that reaches the detector over an integration time $T$ is proportional to the absolute X-ray emission cross section $\sigma_{X-ray}$, and given by:
\begin{equation}
\frac{N_X(p)}{I\ T} = \sigma_{X-ray} \left[B(p)+B_{res}(p) \right], \label{eq:rate}
\end{equation} 
where $I$ is the intensity of the ion beam (ions per second). The residual gas in the collision chamber, when the gas jet is ``on'', contributes via additional counts on the physical peaks through the $B_{res}(p)$ term and is of about 15\% at $p=1.5$~mbar.

Finally $N_X$ is deduced from the number of detected photons $N_{2p}$ and $N_{3p-np}$ of the transitions $1s2p \to 1s^2$ and $1snp \to 1s^2$ with $n\geq3$, taking into account the efficiency $\epsilon$ of the detector for the respective X-ray energy.

For our calculation, we take
\begin{equation}
N_X=\frac {N_{2p}} {\epsilon_{2p}} + \frac {N_{3p-np}} {\epsilon_{3p-np}}.
\end{equation}

\subsection{Extraction of the X-ray cross section values}\label{sec:cross-section}
Measurements are performed with two kinds of target gases: molecular nitrogen and argon.
For both cases systematic studies at different backing pressures $p$ are performed to determine the single collision condition regime. 
\begin{figure}
\begin{center}
\includegraphics[width=0.6\textwidth]{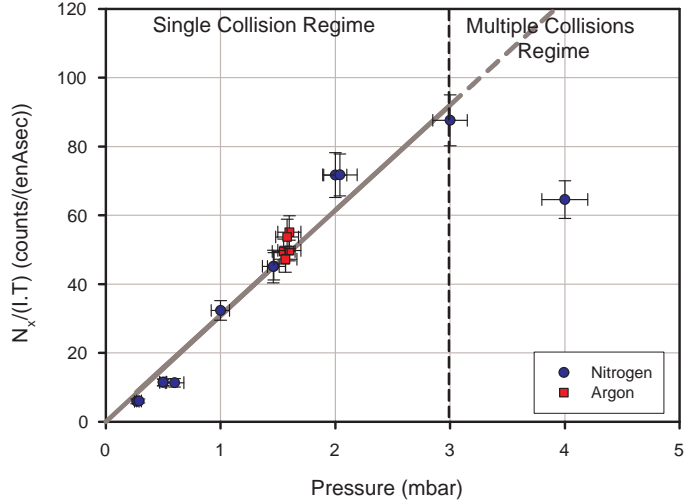}
\caption{(color online) Evolution of the X-ray emission with the baking pressure $p$ of the gas jet, recorded by the SDD at $120^\circ$ for nitrogen and argon targets. The solid grey line corresponds to the expected emission from Eq.~(\ref{eq:rate}).
The observed deviation from this curve at high pressure indicates that multiple-collision start to play a role.}
\label{fig:Nx-p}
\end{center}
\end{figure} 

The evolution of the X-ray rate $N_X(p)/(I\ T)$ is presented in fig.~\ref{fig:Nx-p}. Theoretically \cite{Gray1992}, at low pressure, it displays a linear $p$ dependency and reduces to a pure Poiseuille law ($p^2$ corresponding to viscous flow) at large pressure values. As we can observe, Eq.~(\ref{eq:rate}) reproduces well the experimental data for pressures up to 3~mbar (continuous gray line in the figure), where single collision condition is fulfilled.  
In this region, the data obtained allows us to extract with great precision the absolute value of the X-ray emission cross section, $\sigma_{X-ray} $ (the only adjustable parameter in Eq.~(\ref{eq:rate})), found to be $11.4 \cdot 10^{-15}$~cm$^2$ $\pm 15\%$. 
Furthermore spectra recorded with the two SDDs lead, within the error bars, to the same value. 
It compares well with the unique previous experiment on this system performed by Tawara et al. \cite{Tawara2001}.
It can be noted that the uncertainty has been improved here by more than a factor of two and without referring to any external calibration.

In addition, as seen fig.~\ref{fig:Nx-p}, no noticeable differences, within the error bars, are observed when using argon or molecular nitrogen gaseous targets.
This is easily explained considering the very close ionization potentials of the less bound electrons: -15.8~eV for the argon $3p$ electrons, and -15.5 and -16.8~eV for the N$_2$ $1\pi_u$ and $3\sigma_g$ electrons, respectively.
In those states, 6 electrons are available in both cases.
Moreover, the typical capture radius is around $R_C=7$~\AA  (from classical over-barrier predictions) that is larger than the size of the bi-nitrogen (1~\AA) making molecular nitrogen looking very similar to atomic argon target for the incoming ions. 

For higher pressures, multi-collisions start to take place and as expected, the X-ray rate decreases considerably, since the multiple-excited states which are populated decay mainly by autoionization.
In that respect, and to fulfill the ``single collision condition'', data acquisition was preferentially performed at a backing pressure of around $p=1.5$~mbar to extract reliable results.

\section{Analysis of single and multi-electron capture contribution from  high-resolution X-ray spectra }\label{sec:high-resolution}
\begin{figure}
\begin{center}
\includegraphics[width=\textwidth]{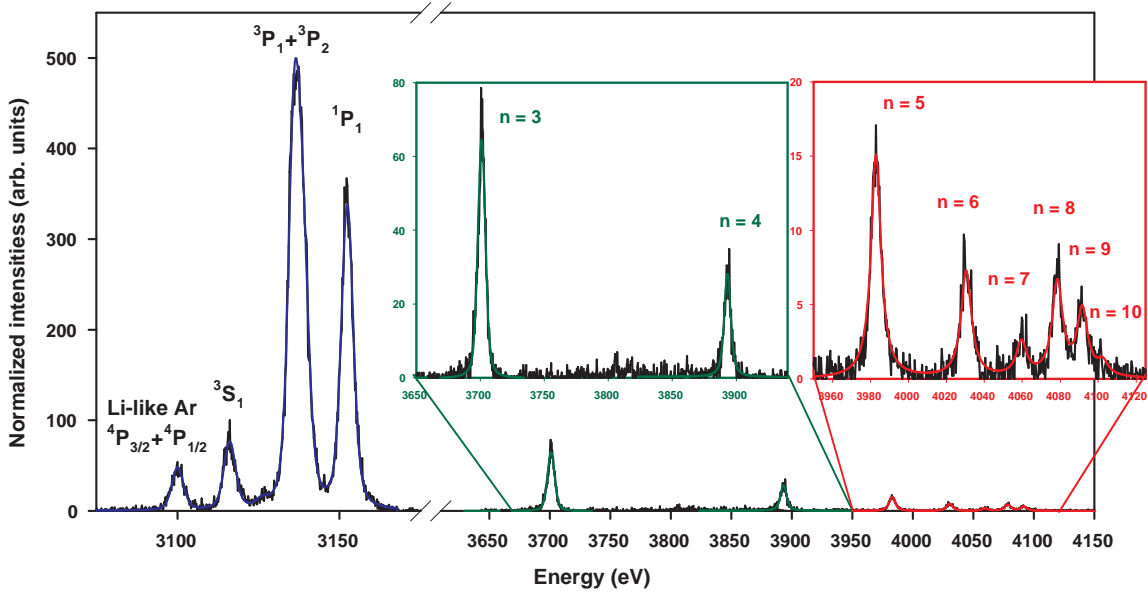}
\caption{(color online) High resolution spectra (background subtracted) of Ar$^{16+}$ X-ray transitions observed with an argon target ($p$=1.5~mbar). 
The He-like Ar $1snp \to 1s^2$ with $n$ up to 10 are visible. For $n=2$, transitions from $1s2p\ ^3\!P_1, ^3\!P_2, ^1\!P_1$ and  $1s2s\ ^3\!S_1$ to the ground level are partially resolved. In addition, M1 $1s2s\ ^3\!S_1 \to 1s^2\ ^1\!S_0$ and Li-like Ar $1s2s2p\ ^4\!P_{1/2,3/2} \to 1s^22s\ ^2\!S_{1/2}$  transitions are observed and well resolved from the $1s2p$ He-like transitions. The continuous colored lines represent the fit using a series of Voigt profiles. Intensities are normalized to the counts recorded by the SDD at $30^\circ$ (see text).}
\label{fig:spectra_ion-gas-hr}
\end{center}
\end{figure}

With the high-resolution spectrometer, an efficient zoom of the SDD spectra is achievable, as shown in figure~\ref{fig:spectra_ion-gas-hr}. A record of the full series of Lyman lines requires to set the spectrometer at a minimum of 3 different Bragg angles.
To ensure sufficient statistics in each peak of the line series, typical acquisition times of a few 1000~s are required for deexcitation of states up to n=4 (i.e., for the $3.0-3.9$~keV energy region), while at least $20~\!000$~s are needed for the end-series region from $n=5$ to 10 (i.e., for the energy range centered around 4.05~ keV).
The whole spectrum, as displayed in figure~\ref{fig:spectra_ion-gas-hr}, exhibits the He-like $1snp\ ^1\!P_1 \to 1s^2\ ^1\!S_0$ transitions extremely well resolved from $n=2$ to 10.
We note that for $n \geq 3$ only the singlet $1snp\ ^1\!P_1$ state decays to the ground state while triplet states decay almost exclusively to the metastable state $1s2s\ ^3\!S_1$ \cite{Savukov2003,Indelicato2005a}.
At low energy, for $n=2$, the He-like $1s2p\ ^3\!P_{1,2} \to 1s^2\ ^1\!S_0$ and $1s2p\ ^1\!P_1 \to 1s^2\ ^1\!S_0$ transitions are observed and well separated. 
M1 $1s2s\ ^3\!S_1 \to 1s^2\ ^1\!S_0$ and Li-like $1s2s2p\ ^4\!P_{1/2,3/2} \to 1s^22s\ ^2\!S_{1/2}$ transitions are also observable and will be discussed in sections~\ref{sec:ratios}.

\subsection{Analysis of the line intensities}
Quantitative values of the transition intensities are obtained from a fit of the lines using a series of Voigt profiles that reproduce well the response function of the crystal spectrometer over the whole energy range studied. Under our experimental conditions, mentioned before, the resolution is primarily driven by the optical quality of the ion beam and the spatial resolution of the detector (i.e., the role of the mosaic spread is negligible). The corresponding width of the Gaussian and Lorentzian terms are kept constant for the energy range explored at a given Bragg angle but their variation, fully under control, is taken into account when varying this angle. In addition, in order to check the reproducibility of the results and to improve the uncertainty on the relative $np$ line intensities, we have also recorded overlap spectra, e.g., transitions from $n=4$ to 7 for instance.

Finally, the background contribution, in terms of intensity and shape has also been precisely determined by recording the same spectra (for each Bragg angle) without the ion beam in the chamber, since, as mentioned before, the background is mainly due to the residual radiation produced by the ion source (see section~\ref{sec:detectors}). Therefore, we have been able to fit the background contribution by a polynomial curve. Its integral over the considered energy range is normalized to the acquisition time. It is worth underlining the importance of such a treatment since the peak-to-background ratio decreases from 30 for  the $1snp\ ^1\!P_1 \to 1s^2\ ^1\!S_0$ transition to 0.3 for the tiny $1s7p \to 1s^2$ transition.

The relative intensity results, extracted from the high-resolution spectra monitored by the SDD detector as previously discussed (see section~\ref{sec:detectors}), are presented in table~\ref{tab:intensities}.
\Table{\label{tab:intensities} For each $n$ value, the $1snp \to 1s^2$ transition intensities in percentage of the total $np$ X-ray emission during the collision of Ar$^{17+}$ ions with Ar and N$_2$ gas target.}
\br
&\centre{2}{Transition intensity (\%)}\\
\ns
Initial   &\crule{2}\\
state $n$ & Nitrogen & Argon\\
\mr
2	& $	88.1	\pm	4.1	$ & $	89.0	 \pm	4.1	$ \\
3	& $	6.6	\pm	1.3	$ & $	6.1	 \pm	1.2	$ \\
4	& $	2.60	\pm	0.58	$ & $	2.15	 \pm	0.48	$ \\
5	& $	1.20	\pm	0.28	$ & $	1.12	 \pm	0.27	$ \\
6	& $	0.56	\pm	0.14	$ & $	0.56	 \pm	0.14	$ \\
7	& $	0.18	\pm	0.05	$ & $	0.18	 \pm	0.05	$ \\
8	& $	0.54	\pm	0.14	$ & $	0.47	 \pm	0.12	$ \\
9	& $	0.27	\pm	0.07	$ & $	0.32	 \pm	0.08	$ \\
10	& $	< 0.02			$ & $	0.05	 \pm	0.02	$ \\
\endTable

As for the low-resolution spectra recorded by the SDD detectors, no significant difference is observed between Ar and N$_2$ targets (see section~\ref{sec:cross-section}).
Therefore we will focus only on the case of argon target for presentation of the detailed analysis and discussions that follow.

\subsection{Analysis of the $n=7-10 \to 1$ transitions: single electron capture}\label{sec:np-1}
Keeping in mind that spectra shown figure~\ref{fig:spectra_ion-gas-hr}, reflect ``only'' the $p$ state populated either through direct capture or by cascades, it clearly comes out that the preferential capture level of the Ar$^{17+}$ ion at low impact velocity (0.53 a.u.) is $n_{pref} \sim  8-9$. This is in good agreement with the simple classical over-barrier model \cite{Ryufuku1980}, which predicts $n_{pref} = 7-9$ for the single-electron capture process from the 3$s$ or 3$p$ Ar states. 
Multi-electron capture populates preferentially multiple excited states lying in lower $n, n', n'' \ldots$ values than single capture \cite {Niehaus1986}. Therefore, assuming that X-ray lines from $n \geq 7$ levels follow almost exclusively single-electron capture, the partial $\{\mathcal{P}_n\}$ of single capture can be extracted directly from the measured transition intensities.
In fact the measured intensities reflect only $1snp\ ^1\!P_1$ states populations decaying to the ground state.
Consequently, to get the population of the $np$ levels, a factor of 4 is introduced, assuming a statistical population between  $^3\!P:^1\!P$ (i.e., corresponding to a ratio of  3:1).
The $\{\mathcal{P}_n\}$ can thus be obtained considering different $\ell$-distributions $\{\mathcal{P}_\ell\}$ population of the sublevels and taking into account the atomic cascade between levels. 
For this purpose,  non-relativistic H-like atom branching ratios are considered \cite{Bethe-Salpeter}.

\begin{figure}
\begin{center}
\includegraphics[width=0.5\textwidth]{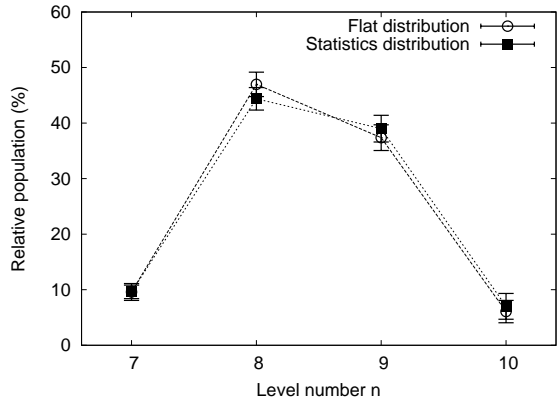}
\caption{ $\{\mathcal{P}_n\}$ distribution of the single-electron capture population for Ar target, assuming different $\ell$-distributions on the $n=7-10$ levels. Note that here $\sum_{n=7}^{10}  \mathcal{P}_n = 100\%$.}
\label{fig:n-distr}
\end{center}
\end{figure}

The results corresponding to the Ar target are presented in figure~\ref{fig:n-distr}.
It is well known that at an ion velocity of 0.53 a.u. the $\ell$-sublevel distribution is most likely of statistical type (\cite{Burgdorfer1986}, or more recently \cite{Kirchner2000,Kirchner2010} for example). Nevertheless to probe the sensibility of the extracted $\{\mathcal{P}_n\}$ on the choice of the $\ell$-distribution, we have assumed two extreme cases: a flat and a statistical distribution.
One can observe that the different assumptions on $\{\mathcal{P}_\ell\}$  do not affect significantly the results, which stay within the error bars. 
We can conclude that the single-electron capture occurs in a reaction window that populates preferentially $n=8-9$ states with a relative narrow distribution in agreement with the Landau-Zener model \cite{Taulbjerg1986} and as observed by Knoop et al. in \cite{Knoop2008}, where the COLTRIM technique for Ar$^{17+}$ on He at  $v=0.4$~a.u has been used.

The accurate knowledge of the relative transmission of the Bragg spectrometer, compared to the SDD detector enables to extract the absolute value of the single-electron capture cross section  $\sigma_{single}$ occurring in ${n=7-10}$. 
Contrary to the $\{\mathcal{P}_n\}$, $\sigma_{single}$ depends strongly on the choice of the $\ell$-distribution. Indeed the extracted cross section value ranges from  $\sigma_{single}=4.6 \cdot10^{-15}$~cm$^2$ (flat) to $12.8 \cdot10^{-15}$~cm$^2$  (statistical). 
The result found can be satisfactorily compared to the value of about $8\cdot10^{-15}$~cm$^2$  obtained by Ali et al. \cite{Ali1994} from coincidence measurements of projectile-target charge exchange for Ar$^{17+}$ on Ar at $v=0.6$~a.u..
Note that those values are also in good agreement with the classical over-barrier model \cite{Janev1985} that predicts $\sigma_{single}$ of the order of $0.5 \pi R_C^2= 6.6\cdot10^{-15}$~cm$^2$.

\subsection{Analysis of the $n=2-6 \to 1$: contribution from multi-electron capture processes}
\begin{figure}
\begin{center}
\includegraphics[width=0.6\textwidth]{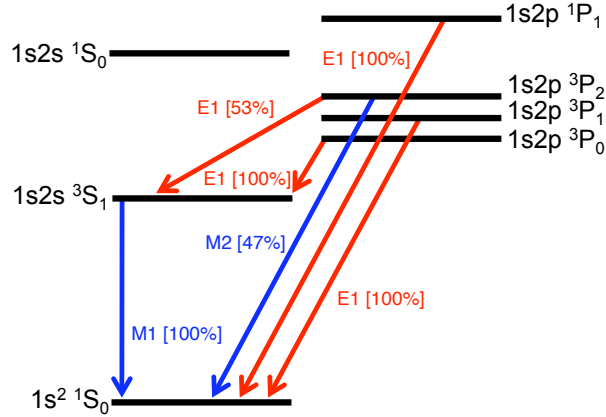}
\caption{(color online) Level scheme of $n=2$ helium-like argon and major one-photon decay branches. The corresponding branching ratio are in bracket \cite{Savukov2003,Indelicato2005a}.}
\label{fig:scheme}
\end{center}
\end{figure}
As mentioned above, multi-electron capture populates preferentially lower levels than the single capture.
These multi-excited states decay preferentially by Auger emission followed by a possible X-ray transition that is indistinguishable from the emission resulting through cascade from single-electron capture. 

\begin{figure}
\begin{center}
\includegraphics[width=0.8\textwidth]{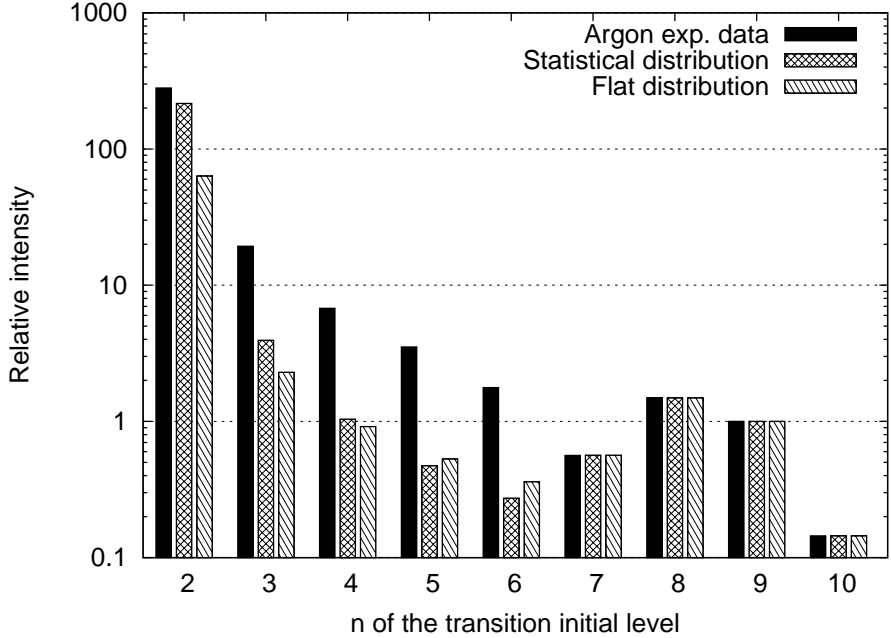}
\caption{Histogram of the $1snp \to 1s^2$ Ar$^{16+}$ transition intensities: Black solid filling for the experimental data normalized with respect to the $1s9p \to 1s^2$ transition for presentation purpose; single and double crossed pattern fillings represent the prediction of the single-electron cascade calculation from the $n=7-10$ levels, assuming a flat or a statistical distribution in the $\ell$-sublevels respectively.}
\label{fig:cascade}
\end{center}
\end{figure}

Once the single capture population probabilities in $n = 7-10$ are extracted, the influence of the multi-electron process can be evaluated.
Considering a given $\ell$-distribution, the population of the different $n\ell$ levels by single capture for $n < 7$ is uniquely determined from the $\{\mathcal{P}_n\}$.
Indeed, the related X-ray emission can be accurately calculated.
In addition to the previously mentioned assumption on the atomic cascade calculation, branching ratios are introduced independently for triplet and singlet states down to $n=2$. For $n=2$, $1s2p \to 1s2s,1s^2$ decays, the complete deexcitation scheme (see fig.~\ref{fig:scheme}) is taken into account, including coupling between triplet and singlet states (i.e., calculations from \cite{Savukov2003,Indelicato2005a}).
The resulting calculated line intensities coming from single capture in $n = 7 - 10$ is presented figure \ref{fig:cascade} for the two extreme $\ell$-distribution (flat and statistical).
Whatever the  $\{\mathcal{P}_\ell\}$, an intensity deficit is observed from $n=3-6$ when comparing the  simulated spectrum to the experimental one. This difference indicates clearly that multi-electron capture contributes significantly starting from the $n=6$ level. 
Moreover, we note that the calculated intensity of the $1s2p \to 1s^2$ transition, which represent 88\% of the total recorded X-ray emission (see table~\ref{tab:intensities}), is very sensitive to the choice of  $\{\mathcal{P}_\ell\}$.
Indeed, in the case of statistical distribution, the high intensity of this transition is explained by the role of the Yrast cascade.
The role of the multi-electron capture can be evaluated for each $\{\mathcal{P}_\ell\}$ and we find that within a flat $\ell$-distribution, the single-electron capture can explain only 22\% of the total X-ray emission while it reaches 71\% for a statistical $\ell$-distribution.

\subsection{Detailed analysis of the $n = 2$ to $n=1$ relative intensities observed with the X-ray spectrometer and relevance of the measured hardness ratio}\label{sec:ratios}
\begin{figure}
\begin{center}
\includegraphics[width=0.8\textwidth]{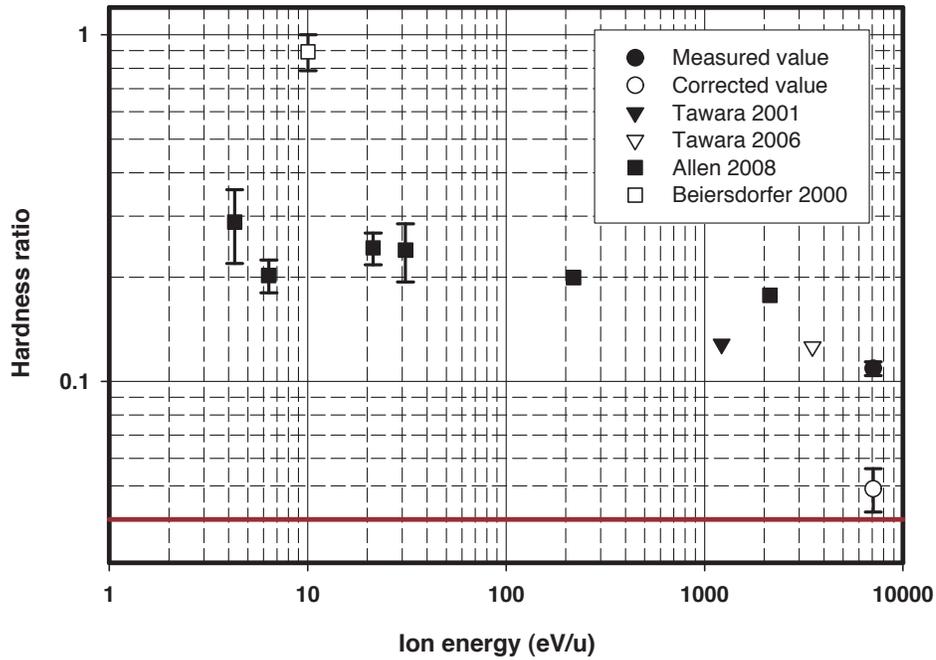}
\caption{(color online) Measured hardness ratio values $\mathcal{H} $ (with one-sigma error bar) for Ar$^{17+} +$ Ar compared with previous experiments (Tawara 2001 \cite{Tawara2001}, Tawara 2006 \cite{Tawara2006}, Allen 2008 \cite{Allen2008}, Beierdorfer 2000 \cite{Beiersdorfer2000}.) The values of Allen 2008 at $E_{ion} = 6.4$~eV/u  and Beierdorfer 2000 correspond to measurements obtained in magnetic trapping experiments. The others are obtained with extracted ion beams.
The empty circle correspond to $\mathcal{H}_{corr}$, the corrected value of $\mathcal{H}$ to take into account the partial detection of the $1s2s\ ^3\!S_1 \to 1s^2\ ^1\!S_0$ M1 transition (see text).
The horizontal solid line corresponds to a $\mathcal{H}$ value expected assuming a statistical population $\{\mathcal{P}_\ell\}$.}
\label{fig:H-ratio}
\end{center}
\end{figure}

The high resolution power of the spectrometer allows for separating different Ar He-like $1s2\ell \to 1s^2$ components as well as He- from Li-like X-ray transitions. 
As observed in figure~\ref{fig:spectra_ion-gas-hr}, emission lines from triplet and singlet $1s2p$ $P$ states can be well distinguished. 
The M1 transition from the metastable state $1s2s\ ^3\!S_1$ to the ground state and the Li-like lines are also well resolved.
The relative intensities between these different X-ray lines deserve a detailed analysis.
First, they allow for checking the consistency of the cascade model used and the estimated contribution of metastable $1s2s$ Ar$^{16+}$ ion beam.
Second, they can be used to refine the reliability on the hardness ratio $\mathcal{H}$ that can be deduced from such measurements. All the values are summarized in table \ref{tab:ratios-n2}.
\Table{\label{tab:ratios-n2} Recorded relative intensities for $n=2 \to 1$ transitions for Ar$^{17+}$ colliding with argon gas. 
$R_{^3\!P/^1\!P}$ refers to the ratio between the He-like Ar $1s2p\ ^3\!P_{1,2} \to 1s^2\ ^1\!S_0$ and $1s2p\ ^1\!P_1 \to 1s^2\ ^1\!S_0$ transitions; 
$R_{^3\!S/(^3P+^1P)}$ and $R^{corr}_{^3\!S/(^3P+^1P)}$  refer to the ratio between the He-like Ar $1s2s\ ^3\!S_1 \to 1s^2\ ^1\!S_0$ and $1s2p\ ^{1,3}\!P_{1,2} \to 1s^2\ ^1\!S_0$ transitions with and without taking into account the correction due the residual vacuum, respectively.
$R_{^4\!P/(^3P+^1P)}$ refers to the ratio between the Li-like $1s2s2p\ ^4\!P \to 1s^22s\ ^2\!S_{1/2}$ and the He-like Ar $1s2s\ ^3\!S_1 \to 1s^2\ ^1\!S_0$ transitions.
Model predictions do not include multi capture.}
\br
& &\centre{2}{Model for inital $\ell$ distribution}\\
\ns
Intenstity &Experimental   &\crule{2}\\
ratio &Values & Statistical & Flat\\
\mr
$R_{^3\!P/^1\!P}$ & $1.91±0.03$	& 1.82	 & 1.91 \\
$R_{^3\!S/(^3P+^1P)}$ &$0.076±0.002$	& 0.040 & 0.053\\
$R^{corr}_{^3\!S/(^3P+^1P)}$ &$0.076±0.002$	& 0.075& 0.101\\
$R_{^4\!P/(^3P+^1P)}$ &$0.048±0.001$	& 0.047 & 0.065 \\
\endTable

a) Experimentally, we find an intensity ratio between  $1s2p\ ^3\!P_{1,2} \to 1s^2\ ^1\!S_0$ and $1s2p\ ^1\!P_1 \to 1s^2\ ^1\!S_0$ transitions equal to $R_{3P/1P} = 1.91\pm0.02$.
Applying the atomic cascade model from the single-electron capture, as described in the previous section, leads to very close values of $R_{3P/1P} = 1.82$ and 1.91 assuming a statistical or flat $\ell$-distribution in the $n=7-10$ levels, respectively.

b) The Li-like argon transition line observed in figure~\ref{fig:spectra_ion-gas-hr} comes from the deexcitation of the $1s2s2p$ states to the ground state.  
Those levels can be produced in the gaseous jet by single electron capture either from the $1s2s\ ^3\!S_1$ Ar$^{16+}$ metastable ions in the incident beam, or by multiple-electron capture from the primary $1s$ Ar$^{17+}$ ion beam. The metastable ion beam contribution can be evaluated taking into account the collision probability on the residual gas in the beam line after the last dipole magnet and before the entrance of the collision chamber as well as within the collision chamber itself before the interaction zone (see section~\ref{sec:setup-det}). It gives rise to a contribution of 10\% that has been indeed confirmed through a measurement of the Lyman X-ray line ratio from Ar$^{16+}$/Ar$^{17+}$ when using Ar$^{18+}$ as incoming ion beam.
Through the capture and cascade processes, the Li-like $1s2sn\ell\ ^4\!L$ and $^2\!L$ levels are populated but X-ray emission from the doublet states are strongly suppressed by auto-ionization during cascade processes \cite{Rohrbein2010}. 
Hence, mostly the $1s2s2p\ ^4\!P_{3/2} + ^4\!P_{1/2} \to 1s^22s\ ^2\!S_{1/2}$ line is emitted as observed on our X-ray spectra, in agreement with previous experiments performed with lighter ions \cite{Strohschein2008,Zouros2008}. 
From the metastable ion contribution in the incoming beam, the intensity ratio between the Li-like $^4\!P$ and the He-like $P$ transitions $R_{4P/(3P+1P)}$ can be estimated around 0.05, which is in good agreement with the experimental value $R^{exp}_{4P/(3P+1P)} = 0.048 \pm 0.001$.
Let us note that X-ray emission from the $^4\!P$ states are fully visible by the spectrometer, contrary to the M1 transition, since they have short lifetime $2-6 \times 10^{-12}$~s \cite{Savukov2003,Costa2001}.

c) The intensity ratio $R_{3S/(3P+1P)}$ between the $1s2s\ ^3\!S_1 \to 1s^2\ ^1\!S_0$ M1 and the $1s2p\ (^3\!P+^1\!P) \to 1s^2\ ^1\!S_0$ transitions can also be predicted from the atomic cascade simulation.
Due to the forbidden transitions from $1snp\ ^3\!P$ states, the $1s2s\ ^3\!S_1$ level is efficiently populated and a value of $R_{3S/(3P+1P)}$ ranging from 0.87 (statistical distribution) to 1.18 (flat distribution) can be expected.
These values appear at first quite far from the experimental measured ratio $R^{exp}_{3S/(3P+1P)}=0.076\pm0.002$. 
This large difference is in fact due to the long lifetime, $0.209~\mu$s, of the $1s2s\ ^3\!S_1$ metastable state \cite{Hubricht1987,Savukov2003,Indelicato2005a}. 
This corresponds, at a projectile velocity of 0.53~a.u., to a decay length of 24.5~cm while a spatial window of only $\sim 2$~cm around the center of the gaseous jet is viewed by the X-ray spectrometer. 
As a result, only about 4.5\% of the total M1 emission is detected compared to 100\% for the $1s2p \to 1s^2$ transitions. An intensity ratio $R_{3S/(3P+1P)}$ ranging from 0.040 (statistical distribution) to 0.053 (flat distribution) is thus expected.
When the capture from the residual vacuum is included, as discussed previously, such values go up to 0.075 (statistical distribution)  and 0.101 (flat distribution), which is in good agreement with the observed ratio, keeping in mind that the contribution of multi-electron capture is neglected.

d) In the case of He-like transitions, the relative long lifetime of $1s2s\ ^3\!S_1$ M1 level has also important consequences on the relevance of the measured hardness ratio $\mathcal{H}$, as already pointed out by Tawara et al. \cite{Tawara2001}. Indeed, transposition of the $\mathcal{H}$ value measured via ``laboratory ion-atom collisions'' towards interpretation of spectra from comets and solar wind should be made with caution.
We remind that the hardness ratio $\mathcal{H}$ is the ratio between the unresolved $n>2 \to 1$ and the $2 \to 1$ transitions. From our experiment, the measured $\mathcal{H}$ values are found to be $0.106\pm0.005$ and $0.109\pm0.005$ with the two SDD detectors at 120 and 30$^\circ$, consistent with previous measurements performed in the same experimental conditions as shown in fig.~\ref{fig:H-ratio}. 
Nevertheless, great care has to be taken when dealing with X-ray spectra observed from comet or solar wind for which 100\% of the He-like M1 transition is visible.
With our experiment, we have recorded only a fraction (4.5\% as mentioned above) of this transition.
Consequently, we have to correct our measured data by a factor of $0.45\pm0.06$ leading to a hardness ratio $\mathcal{H}_{corr}= 0.049 \pm 0.007$.
When considering ions of different Z value than argon, e.g. C, N, O, Ne, Mg and Si, which have been identified at the origin of the X-ray emissions from interstellar medium \cite{Krasnopolsky2004}, attention has to be payed since the correction factor depends on the metastable state lifetime in addition of the dependency on the collision velocity (through the decay length and the $\{\mathcal{P}_\ell\}$).

Finally, in this respect also, comparing measurements performed with extracted ions of typically a few keV/u and trapped ions as in EBIT experiments is not trivial.
As observed in figure~\ref{fig:H-ratio}, for Ar$^{17+}$ incident beam, the corrected $\mathcal{H}_{corr}$ value we found reaches almost the statistical limit as expected at the relative high-velocity of the slow collision energy regime.

\section{Conclusions}
Using low- and high-resolution spectroscopy, projectile X-ray emission is studied for (quasi) symmetric heavy collision systems at low energy, where Ar$^{17+}$ ions collide with Ar and N$_2$ targets at $v=0.53$ a.u..
The determination of the ion beam-gaseous jet overlap, coupled to a perfect knowledge of our SDD detectors efficiencies, allow us to extract the absolute X-ray emission cross section, found to be  $11.4 \cdot 10^{-15}$~cm$^2$ $\pm 15\%$ in agreement with the only previous experiment on this system performed by Tawara et al. \cite{Tawara2001} but with an improvement in the experimental uncertainty by more than a factor of 2.
Moreover, with our direct method of measurement, contrary to ref.~\cite{Tawara2001}, no reference cross section is needed.

Differently from experiments where low-resolution X-ray spectroscopy were employed \cite{Beiersdorfer2000, Allen2008}, the high resolving power of the crystal spectrometer enables to have access to additional information on single and multi-electron processes without applying coincidence with target charge state detection.
In our case, we are able to resolve the whole He-like Ar$^{16+}$ Lyman series from $n=2$ to 10 as well as the fine structure of $n=2\to1$ transitions. For each $n$ value, the relative $1snp \to 1s^2$ intensity are obtained showing that the $1s2p \to 1s^2$ transition represent almost 90\% of the total X-ray emission whatever the gaseous target (Ar or N$_2$ ).
From the analysis of the $n=7-10 \to 1$ line intensities, the preferential $n_{pref}$ is precisely determined to be $8-9$.
Moreover, under the hypothesis that such transitions are exclusively due to the single capture, the relative distribution $\{\mathcal{P}_n\}$ is extracted.
In particular, we demonstrate this $\{\mathcal{P}_n\}$ to be independent on the $\ell$-sublevel distribution population $\{\mathcal{P}_\ell \}$.
Finally, an evaluation of the single-electron capture cross section can be also extracted.
In this case, the result is influenced by the choice of $\{\mathcal{P}_\ell \}$ and found to be between $\sigma_{single}=4.6 \cdot10^{-15}$~cm$^2$ (flat distribution) and $12.8 \cdot10^{-15}$~cm$^2$ (statistical distribution). 
With the knowledge of $\{\mathcal{P}_n\}$, the multi-electron capture can be also evaluated through the comparison between the experimental data and the expected intensities of the $n=2-6 \to 1$ lines from single-electron capture using a cascade code.
Quantitatively, we find that the single-electron capture can explain only 22\% of the total X-ray emission with a flat $\{\mathcal{P}_\ell \}$ while it reaches 71\% for a statistical  $\{\mathcal{P}_\ell \}$. Those findings provide a stringent test on the consistency of a given theoretical description able to account for single and multiple capture processes in such symmetric systems when it will be available.

Finally, the analysis of the intensities of the different He-like $1s2\ell \to 1s^2$ transitions (involving the $1s2p\ ^1\!P$ and $^3\!P$ states and the metastable $1s2s\ ^3\!S_1$ state) as well as the Li-like $1s2s2p\ ^4\!P \to 1s^22s$ transitions, allow  to control the contamination from metastable Ar$^{16+}$ ions in the incoming ion beam. 
The consistency of our cascade calculations is checked since all the evaluated relative peak intensities are found in rather good agreement with the experimental ratios. 
This fine analysis gives argues to discuss on the hardness ratio $\mathcal{H}$, commonly used to analyze low-resolution X-ray spectra. 
In particular, we point out that  special care has to be taken when experiments on the same collision system but with different setups are compared.
Indeed, the partial detection of the $1s2s\ ^3\!S_1$ metastable state decay can drastically affect the value of the hardness ratio, by a factor of two in the case presented in this paper. 

In conclusion, it is worth mentioning that accurate determination of $n_{pref}$ and $\{\mathcal{P}_n\}$ through X-ray measurements are achievable only with a high-resolution spectrum, where the different transitions can be well separated.
Due to the dependency on the $\{\mathcal{P}_\ell\}$ on the single capture process, only rough values can be obtained for single capture cross section and multiple capture contribution in X-ray spectra.
However, high-resolution spectroscopy can be advantageously used to test the consistency of a full calculation of the different processes involved in such symmetric collisions.
The achievement of a more quantitative estimation requires to apply coincidence techniques between  high-resolution X-ray spectroscopy and ion charge state detection, or at least to perform simultaneously COLTRIM type experiments, which could be done in the future.

\ack
The authors would like to thank S.Bari, H.M.~Dang, S.~Geyer, H.~Merabet, T.~Schlath\"olter for their participation in the data acquisition. We also like to acknowledge the support given by the GANIL and CIMAP engineers and technical staff. In particular we are grateful to L.~Maunoury, J-Y.~Pacquet and J-M.Ramillon  for providing us their expertise. We also want to thank 
T.~Kirchner for very fruitful discussions on theoretical aspects.
This experiment is supported by a grant from ``Agence Nationale pour la Recherche (ANR)'' number \emph{ANR-06-BLAN-0223} and Helmholtz Alliance HA216/EMMI. 
L'Institut des Nanosciences de Paris (INSP) is ``Unit\'e Mixte de Recherche du CNRS et de l'UPMC n$^{\circ}$ 7588''.

\section*{References}
\bibliographystyle{unsrt}

\begin{thebibliography}{}

\end{thebibliography}


\begin{thebibliography}{10}

\bibitem{Tawara2001}
H.~Tawara, P.~Richard, U.~I. Safronova, and P.~C. Stancil.
\newblock K x-ray production in {H}-like {Si}$^{13+}$, {S}$^{15+}$, and
  {Ar}$^{17+}$ ions colliding with various atom and molecule gas targets at low
  collision energies.
\newblock {\em Phys. Rev. A}, 64(4):042712, 2001.

\bibitem{Barat1992}
M.~Barat and P.~Roncin.
\newblock Multiple electron capture by highly charged ions at kev energies.
\newblock {\em J. Phys. B}, 25(10):2205, 1992.

\bibitem{Mack1989}
M.~Mack, J.~H. Nijland, P.~v~d Straten, A.~Niehaus, and R.~Morgenstern.
\newblock Correlation in double electron capture in collisions of fully
  stripped ions on {He} and {H}$_{2}$.
\newblock {\em Phys. Rev. A}, 39(8):3846, 1989.

\bibitem{Vernhet1988}
D.~Vernhet, A.~Chetioui, J.~P. Rozet, C.~Stephan, K.~Wohrer, A.~Touati, M.~F.
  Politis, P.~Bouisset, D.~Hitz, and S.~Dousson.
\newblock Characteristics of single capture nl distributions and double capture
  probabilities in slow collisions of {$Al^{13+}$}, {$Al^{12+}$} and
  {$Ne^{9+}$} ions with two-electron targets ({$He$}, {$H_2$}).
\newblock {\em J. Phys. B}, 21(23):3949, 1988.

\bibitem{Chetioui1990}
A.~Chetioui, F.~Martin, J.~P. Rozet, A.~Touati, L.~Blumendeld, D.~Vernhet,
  K.~Wohrer, C.~Stephan, M.~Barat, M.~N. Gbouriaud, H.~Laurent, and P.~Roncin.
\newblock Doubly excited states populated in collisions of {O}$^{8+}$ ions with
  {He} and {H}$_2$ at 1.24 {keV} amu$^{-1}$.
\newblock {\em J. Phys. B}, 23(20):3659, 1990.

\bibitem{Stolterfoht1997}
Nikolaus Stolterfoht, Robert~D. DuBois, and Roberto~D. Rivarola.
\newblock {\em The Physics of Multiply and Highly Charged Ions: Interactions
  with matter}.
\newblock Springer Series on Atoms and Plasmas. Springer, 1997.

\bibitem{Currel2003}
Fred~J. Currel.
\newblock {\em The Physics of Multiply and Highly Charged Ions: Interactions
  with matter}.
\newblock Kluwer Academic Publisher, 2003.

\bibitem{Ryufuku1980}
Hiroshi Ryufuku, Ken Sasaki, and Tsutomu Watanabe.
\newblock Oscillatory behavior of charge transfer cross sections as a function
  of the charge of projectiles in low-energy collisions.
\newblock {\em Phys. Rev. A}, 21(3):745, 1980.

\bibitem{Taulbjerg1986}
K.~Taulbjerg.
\newblock Reaction windows for electron capture by highly charged ions.
\newblock {\em J. Phys. B}, 19(10):L367, 1986.

\bibitem{Fritsch1984}
Wolfgang Fritsch and C.~D. Lin.
\newblock Atomic-orbital-expansion studies of electron transfer in bare-nucleus
  ${Z} ({Z}=2,4-8)$ hydrogen-atom collisions.
\newblock {\em Phys. Rev. A}, 29(6):3039--3051, 1984.

\bibitem{Winter1984}
Thomas~G. Winter and C.~D. Lin.
\newblock Triple-center treatment of electron transfer and excitation in p-{H}
  collisions.
\newblock {\em Phys. Rev. A}, 29(2):567--582, 1984.

\bibitem{Kimura1987}
M.~Kimura and C.~D.Lin.
\newblock A unified atomic-orbital and molecular- orbital matching method for
  ion-atom and atom-atom collisions.
\newblock {\em Comments on Atomic and Mol. Phys.}, 20(35), 1987.

\bibitem{Errea1987}
L.~F. Errea, J.~M. Gomez-Llorente, L.~Mendez, and A.~Riera.
\newblock Convergence study of ${He}^{2+} +{H}$ and ${He}^+ +{H}^+$ charge
  exchange cross sections using a molecular approach with an optimised common
  translation factor.
\newblock {\em J. Phys. B}, 20(22):6089, 1987.

\bibitem{Knoop2008}
S.~Knoop and et~al.
\newblock Single-electron capture in kev {Ar}$^{15+...18+}$ + {He} collisions.
\newblock {\em J. Phys. B}, 41(19):195203, 2008.

\bibitem{Barany1985}
A.~Barany, G.~Astner, H.~Cederquist, H.~Danared, S.~Huldt, P.~Hvelplund,
  A.~Johnson, H.~Knudsen, L.~Liljeby, and K.~G. Rensfelt.
\newblock Absolute cross sections for multi-electron processes in low energy
  {$Ar^{q+}-Ar$} collisions: Comparison with theory.
\newblock {\em Nucl. Instrum. Meth. Phys. Res. B}, 9(4):397--399, 1985.

\bibitem{Ali1994}
R.~Ali, C.~L. Cocke, M.~L.~A. Raphaelian, and M.~Stockli.
\newblock Multielectron processes in 10-{keV}/u {Ar}$^{q+}$ ($5 \leq q \leq
  17$) on {Ar} collisions.
\newblock {\em Phys. Rev. A}, 49(5):3586, 1994.

\bibitem{Beiersdorfer2000}
P.~Beiersdorfer, R.~E. Olson, G.~V. Brown, H.~Chen, C.~L. Harris, P.~A. Neill,
  L.~Schweikhard, S.~B. Utter, and K.~Widmann.
\newblock X-ray emission following low-energy charge exchange collisions of
  highly charged ions.
\newblock {\em Phys. Rev. Lett.}, 85(24):5090, 2000.

\bibitem{Tawara2006}
Hiro Tawara, Endre Takacs, Tibor Suta, Karoly Makonyi, L.~P. Ratliff, and J.~D.
  Gillaspy.
\newblock K x rays produced in collisions of bare ions with atoms: Contribution
  of multiple-electron transfer in ${Kr}^{36+}$, ${Ar}^{18+}$, and ${Ne}^{10+}
  + {Ar}$ collisions.
\newblock {\em Phys. Rev. A}, 73(1):012704--5, 2006.

\bibitem{Allen2008}
F.~I. Allen, C.~Biedermann, R.~Radtke, G.~Fussmann, and S.~Fritzsche.
\newblock Energy dependence of angular momentum capture states in charge
  exchange collisions between slow highly charged argon ions and argon
  neutrals.
\newblock {\em Phys. Rev. A}, 78(3):032705--7, 2008.

\bibitem{Cravens2002}
T.~E. Cravens.
\newblock X-ray emission from comets.
\newblock {\em Science}, 296(5570):1042--1045, 2002.

\bibitem{Krasnopolsky2004}
Vladimir~A. Krasnopolsky, Jason~B. Greenwood, and Philip~C. Stancil.
\newblock X-ray and extreme ultraviolet emissions from comets.
\newblock {\em Space Sci. Rev.}, 113(3):271--373, 2004.

\bibitem{Otranto2011}
S.~Otranto and R.~E. Olson.
\newblock X-ray emission cross sections following {Ar}$^{18+}$ charge-exchange
  collisions on neutral argon: The role of the multiple electron capture.
\newblock {\em Phys. Rev. A}, 83(3):032710, 2011.

\bibitem{Maunoury2002}
L.~Maunoury, R.~Leroy, T.~Been, G.~Gaubert, L.~Guillaume, D.~Leclerc,
  A.~Lepoutre, V.~Mouton, J.~Y. Pacquet, J.~M. Ramillon, and R.~Vicquelin.
\newblock {LIMBE}: A new facility for low energy beams.
\newblock {\em Rev. Sci. Instrum.}, 73(2):561--563, 2002.

\bibitem{Prigent2009}
C.~Prigent, E.~Lamour, J.~Merot, B.~Pascal, J.~P. Rozet, M.~Trassinelli,
  D.~Vernhet, J.~Y. Pacquet, L.~Maunoury, F.~Noury, and J.~M. Ramillon.
\newblock X-ray spectroscopy characterization of {Ar}$^{17+}$ produced by an
  ecris in the afterglow mode.
\newblock {\em J. Phys. CS}, 163:012111, 2009.

\bibitem{Giordmaine1960}
J.~A. Giordmaine and T.~C. Wang.
\newblock Molecular beam formation by long parallel tubes.
\newblock {\em J. Appl. Phys.}, 31(3):463--471, 1960.

\bibitem{Gray1992}
D.~C. Gray and H.~H. Sawin.
\newblock Design considerations for high-flux collisionally opaque
  molecular-beams.
\newblock {\em J. Vac. Sci. Technol. A}, 10(5):3229--3238, 1992.

\bibitem{Rugamas2000}
F.~Rugamas, D~Roundy, G~Mikaelian, G~Vitug, M~Rudner, J~Shih, D~Smith,
  J~Segura, and M~A Khakoo.
\newblock Angular profiles of molecular beams from effusive tube sources: I.
  experiment.
\newblock {\em Meas. Sci. Technol.}, 11(12):1750, 2000.

\bibitem{Lamour2009}
Emily Lamour, Christophe Prigent, Benjamin Eberhardt, Jean~Pierre Rozet, and
  Dominique Vernhet.
\newblock {2E1} ${Ar}^{17+}$ decay and conventional radioactive sources to
  determine efficiency of semiconductor detectors.
\newblock {\em Rev. Sci. Instrum.}, 80(2):023103--7, 2009.

\bibitem{Rozet1985}
J.~P. Rozet, P.~Chevallier, P.~Legagneux-Piquemal, A.~Chetioui, and C.~Stephan.
\newblock Capture cross sections in highly excited p states of {Ar}$^{17+}$ in
  high-velocity collisions of 250 {MeV} {A}r$^{18+}$ on {N}.
\newblock {\em J. Phys. B}, 18(5):943, 1985.

\bibitem{Vernhet1998}
D.~Vernhet, J.~P. Rozet, I.~Bailly-Despiney, C.~Stephan, A.~Cassimi, J.~P.
  Grandin, and L.~J. Dubé.
\newblock Observation of dynamical substate mixing of fast ions in solids.
\newblock {\em J. Phys. B}, 31(1):117, 1998.

\bibitem{Gumberidze2010}
A.~Gumberidze, M.~Trassinelli, N.~Adrouche, C.~I. Szabo, P.~Indelicato,
  F.~Haranger, J.~M. Isac, E.~Lamour, E.~O. Le~Bigot, J.~Merot, C.~Prigent,
  J.~P. Rozet, and D.~Vernhet.
\newblock Electronic temperatures, densities, and plasma x-ray emission of a
  14.5 {GHz} electron-cyclotron resonance ion source.
\newblock {\em Rev. Sci. Instrum.}, 81(3):033303--10, 2010.

\bibitem{Prigent2005}
C.~Prigent.
\newblock {\em L'Émission X : un Outil et une Sonde pour l'Interaction
  Laser--Agrégats}.
\newblock PhD thesis, Université Pierre et Marie Curie, Paris, 2004.

\bibitem{Wohrer2000}
K.~Wohrer, M.~Chabot, R.~Fossé, and D.~Gardès.
\newblock A method for ``on-line'' determination of beam-jet overlaps;
  application to cluster fragmentation studies.
\newblock {\em Rev. Sci. Instrum.}, 71:2025--2032, 2000.

\bibitem{Gigante1989}
G.~E. Gigante and A.~L. Hanson.
\newblock The assessment of geometrical effects on {C}ompton profile
  measurements.
\newblock {\em Nucl. Instrum. Meth. Phys. Res. A}, 280(2-3):299--300, 1989.

\bibitem{Conway2010}
John~T. Conway.
\newblock Analytical solution for the solid angle subtended at any point by an
  ellipse via a point source radiation vector potential.
\newblock {\em Nucl. Instrum. Meth. Phys. Res. A}, 614(1):17--27, 2010.

\bibitem{Savukov2003}
I.~M. Savukov, W.~R. Johnson, and U.~I. Safronova.
\newblock Multipole ({E1}, {M1}, {E2}, {M2}) transition wavelengths and rates
  between states with n<=6 in helium-like carbon, nitrogen, oxygen, neon,
  silicon, and argon.
\newblock {\em At. Data Nucl. Data Tables}, 85(1):83--167, 2003.

\bibitem{Indelicato2005a}
P.~Indelicato and J.P. Desclaux.
\newblock {MCDFGME}, a {MultiConfiguration Dirac Fock} and general matrix
  elements program (release 2005).
\newblock \url{http://dirac.spectro.jussieu.fr/mcdf}, 2005.

\bibitem{Niehaus1986}
A.~Niehaus.
\newblock A classical model for multiple-electron capture in slow collisions of
  highly charged ions with atoms.
\newblock {\em J. Phys. B}, 19(18):2925--2937, 1986.

\bibitem{Bethe-Salpeter}
H.~B. Bethe and E.~E. Salpeter.
\newblock {\em Quantum Mechanics of One- and Two-Electron Atoms}.
\newblock Springer-Verlag, first edition, 1957.

\bibitem{Burgdorfer1986}
J.~Burgdorfer, R.~Morgenstern, and A.~Niehaus.
\newblock Angular momentum distribution in the classical over-barrier model for
  electron capture into highly charged slow projectiles.
\newblock {\em J. Phys. B}, 19(14):L507--L513, 1986.

\bibitem{Kirchner2000}
T.~Kirchner, M.~Horbatsch, H.~J. Ludde, and R.~M. Dreizler.
\newblock Time-dependent screening effects in ion-atom collisions with many
  active electrons.
\newblock {\em Phys. Rev. A}, 62(4):042704, 2000.

\bibitem{Kirchner2010}
T.~Kirchner.
\newblock private communication, 2011.

\bibitem{Janev1985}
R.~K. Janev and Hannspeter Winter.
\newblock State-selective electron capture in atom-highly charged ion
  collisions.
\newblock {\em Phys. Rep.}, 117(5-6):265--387, 1985.

\bibitem{Rohrbein2010}
D.~R\"ohrbein, T.~Kirchner, and S.~Fritzsche.
\newblock Role of cascade and auger effects in the enhanced population of the
  ${C}^{3+}(1s2s2p\ ^{4}\!{P})$ states following single-electron capture in
  ${C}^{4+}(1s2s\ ^{3}\!{S})-{He}$ collisions.
\newblock {\em Phys. Rev. A}, 81(4):042701, 2010.

\bibitem{Strohschein2008}
D.~Strohschein, D.~R\"ohrbein, T.~Kirchner, S.~Fritzsche, J.~Baran, and J.~A.
  Tanis.
\newblock Nonstatistical enhancement of the $1s2s2p\ \!^{4}{P}$ state in
  electron transfer in 0.5--1.0-{MeV}/u ${C}^{4,5+} + {He}$ and ${Ne}$
  collisions.
\newblock {\em Phys. Rev. A}, 77(2):022706, 2008.

\bibitem{Zouros2008}
T.~J.~M. Zouros, B.~Sulik, L.~Gulyas, and K.~Tokesi.
\newblock Selective enhancement of $1s2s2p^{4}{P}_{J}$ metastable states
  populated by cascades in single-electron transfer collisions of
  ${F}^{7+}(1s^{2}/1s2s\ ^{3}\!{S})$ ions with ${He}$ and ${H}_{2}$ targets.
\newblock {\em Phys. Rev. A}, 77(5):050701, 2008.

\bibitem{Costa2001}
A.~M. Costa, M.~C. Martins, F.~Parente, J.~P. Santos, and P~Indelicato.
\newblock Dirac-fock transition energies and radiative and radiationless
  transition probabilities for {$Ar^{9+}$} to {$Ar^{16+}$} ion levels with
  k-shell holes.
\newblock {\em At. Data Nucl. Data Tables}, 79(2):223--39, 2001.

\bibitem{Hubricht1987}
G.~Hubricht and E.~Träbert.
\newblock The ${Ar}^{16+} 2\ \!^3{S}_1$ lifetime from a measurement on recoil
  ions.
\newblock {\em Z. f. Physik D}, 7(3):243--250, 1987.

\end{thebibliography}

\end{document}